\def\year{2020}\relax
\documentclass[letterpaper]{article} 
\usepackage{aaai20}  
\usepackage{times}  
\usepackage{helvet} 
\usepackage{courier}  
\usepackage[hyphens]{url}  
\usepackage{graphicx} 
\usepackage{graphicx}  
\usepackage{makecell}
\nocopyright
\usepackage{hyperref}

\title{\thetitle}

\author{Chamin Hewa Koneputugodage,\textsuperscript{\rm 1} Rhys Healy,\textsuperscript{\rm 1} Sean Lamont,\textsuperscript{\rm 1} Ian Mallett,\textsuperscript{\rm 1}\\ \Large \textbf{Matt Brown,\textsuperscript{\rm 1} Matt Walters,\textsuperscript{\rm 1} Ushini Attanayake,\textsuperscript{\rm 1} Libo Zhang,\textsuperscript{\rm 1,2}} \\ \Large \textbf{Roger T. Dean,\textsuperscript{\rm 3} Alexander Hunter,\textsuperscript{\rm 1} Charles Gretton,\textsuperscript{\rm 1} Christian Walder,\textsuperscript{\rm 1,4}}\\
\textsuperscript{\rm 1}Australian National University\\
\textsuperscript{\rm 2}Nanjing University\\
\textsuperscript{\rm 3}The MARCS Institute for Brain, Behaviour and Development, Western Sydney University\\
\textsuperscript{\rm 4}Data61, CSIRO\\ 
\{first.last\}@\{anu.edu.au, westernsydney.edu.au, smail.nju.edu.cn\}\\
}

\newcommand{\citet}[1]{\citeauthor{#1} \shortcite{#1}}
\newcommand{\citep}{\cite}

\usepackage{amsmath}
\usepackage{amssymb}
\usepackage{algorithm}
\usepackage[noend]{algpseudocode}
\usepackage{bm}
\usepackage{bbm}
\usepackage[usenames,dvipsnames,svgnames,table]{xcolor}
\usepackage{slashbox}
\usepackage{mathtools}
\mathtoolsset{showonlyrefs}
\usepackage{listings}

\usepackage{multirow}
\usepackage{caption}
\usepackage{subcaption}
\usepackage{psfrag}

\DeclareMathOperator*{\bernoulli}{Bernoulli}
\DeclareMathOperator*{\true}{True}
\DeclareMathOperator*{\false}{False}
\DeclareMathOperator*{\truet}{T}
\DeclareMathOperator*{\falsef}{F}

\newcommand{\ts}{^{(s)}}

\newcommand{\alls}{\ensuremath{1,2,\dots,S}}
\newcommand{\rtext}[1]{\hfill \textcolor{gray}{#1}}

\newcommand{\tabincell}[2]{\begin{tabular}{@{}#1@{}}#2\end{tabular}}
\newcommand{\sref}[2]{#1~\ref{#2}}
\newcommand{\thetitle}{Computer Assisted Composition in Continuous Time}
\title{\thetitle}

\begin{document}

\maketitle

\begin{abstract}
We address the problem of combining sequence models of symbolic music with user defined constraints. For typical models this is non-trivial as only the conditional distribution of each symbol given the earlier symbols is available, while the constraints correspond to arbitrary times. Previously this has been addressed by assuming a discrete time model of fixed rhythm. We generalise to continuous time and arbitrary rhythm by introducing a simple, novel, and efficient particle filter scheme, applicable to general continuous time point processes. Extensive experimental evaluations demonstrate that in comparison with a more traditional beam search baseline, the particle filter exhibits superior statistical properties and yields more agreeable results in an extensive human listening test experiment.
\end{abstract}


\section{Introduction}

The recent popularity of deep learning methods has given rise to progress in computational models and algorithms for music generation \cite{aimusicsurvey,nierhaus,Briot2017DeepLT}. While fully automated systems are interesting in their own right, many researchers strive to combine machine learning models with human input \citep{jamfactory,continuator,flowcomposer,magentastudio}. 

Of the various approaches to computer assisted music composition, a number of broad categories are relevant to understanding the nature of the present contribution. Perhaps the simplest scheme is the ``continuator'', which allows the user to provide a musical fragment that is then extended forward in time by the algorithm \cite{continuator,performancernn}. Assuming that we have the usual causally factorised model (that is, we have access to the conditional $p(x_t|x_1,x_2,\dots,x_{t-1})$), these approaches sample the remainder of a piece given its beginning. However, the continuator approach is restrictive in that it may only condition on the start of the piece. This can be overcome by two distinct families of alternative approaches.

The first family of approaches treats the completion of the musical piece as an \textit{infilling} problem \citep{infillingpiano,musictransformer}. Here the input is replaced with random masks, and a model is learnt to fill in the masked segments. While interesting and powerful, this approach conflates the modelling problem with the subsequent composition step. As a result, it is unclear how best to train such a model. 

The second general approach (to which this paper belongs) combines a typical causal model of music with user defined constraints by introducing a subsequent optimisation or conditioned sampling scheme. Pieces are subsequently generated to satisfy these constraints while being representative of the learned music model \cite{exactmarkovbp,pfrnn}. This has the advantage of permitting standard models of (unconstrained) music, while placing the burden of constraining the model on the composition algorithm. The challenging nature of this subsequent step has meant that only relatively simple models, typically with discrete time and fixed rhythmic structures, may be considered. For example, limiting the Markov blanket of a given musical event to a small temporal window surrounding it permits simple Gibbs sampling techniques \cite{deepbach}, but precludes the modelling of long range structure using advanced neural sequence models.

In this paper we present the first generic scheme for combining user constraints with a general continuous time point process model \citep{baddeleypps} of music which enjoys the usual convenient causal factorisation structure, and which has no  limitation on the temporal range of the dependencies modeled. We formulate the problem in \sref{Section}{sec:formulation}, before introducing our main contribution, a novel particle filter scheme \citep{doucet2009tutorial}, in \sref{Section}{sec:pf}. \sref{Section}{SMC_music} explains how we apply the generic particle filter scheme to the problem of constrained music generation, \sref{Section}{sec:results} details extensive quantitative experiments, and \sref{Section}{sec:conclusion} concludes the paper.

\section{Problem Formulation}
\label{sec:formulation}
In the following sub-sections we first introduce some necessary notation and definitions, and then formally define the conditional sampling problem designed for generating computer assisted music in continuous time.
\subsection{Notation and Basic Definitions}
\subparagraph{Notation}
Bold such as $\bm{x}$ denotes a vector (or sequence) whose $i$-th element we denote $x_i$. $\mathbb{R}^*=\bigcup_{i=0}^\infty \mathbb{R}^i$ is the set of arbitrary length sequences, including the empty sequence. $\ell$ returns the length of its argument, so that $\bm{x} \in \mathbb{R}^d \Rightarrow \ell(\bm{x}) = d$. We also use $x_\ell$ to denote the last element of $\bm{x}$, and $x_{\ell-1}$ the second last. We combine this with slices such as $\bm{x}_{<i}$, $\bm{x}_{1:i-1}$ and $\bm{x}_{<\ell}$, which have the usual meaning (first two being equivalent, and the third being a sequence with all but the last element of $\bm{x}$). Finally, in the algorithm pseudo-code we use \textit{e.g.} $1:S$ to indicate iteration over $1,2,\dots,S$.
\subparagraph{Point Process Formulation}
Suppose we have a general one dimensional point process \citep{baddeleypps} defined by
\begin{align}
x_0 & = 0 \\
\label{eqn:xi}
x_i & \sim x_{i-1} + D(\bm{x}_{<i}), ~~~ i \in \mathbb{N}_+,
\end{align}
where $\bm{x}_{<i} = ( x_0, x_1, \dots , x_{i-1})^\top \in \mathbb{R}^i$ and 
\begin{align}
D: \mathbb{R}^* \rightarrow \Delta,
\end{align}
where $\Delta$ is the set of random variables with samples in $(0, +\infty]$. Hence, $D$ is a function which maps vectors (of the history of the sequence) to inter-arrival distributions (for the next sample in the sequence).
Let the random variables $\bm{x}_{<i}$ generated in this way be distributed as
\begin{align}
\bm{x}_{<i} \sim X(D,i).
\end{align}
This means that $X(D, i)$ is a random variable parametrised by $D$ and $i$, with samples lying in $\mathbb{R}^i$. These samples are non-decreasing sequences with first element equal to zero. The probability of a sequence $\bm{x}_{1:i}$ can be written
in terms of the conditional probability density function (p.d.f.) as
\begin{align}
p(\bm{x}_{1:i}) &\propto \prod_{k=1}^{i}f(x_{k}|\bm{x}_{<k}). \label{factorised}
\end{align}
\subparagraph{Restriction Operator}
Define the operator $\Pi_\mathcal{I}$ which restricts sequences to the set $\mathcal{I} \subset \mathbb{R}$:
\begin{align}
\Pi_\mathcal{I}: \, & \mathbb{R}^* \rightarrow \mathbb{R}^* \\
& \bm{x} \mapsto \Pi_\mathcal{I}(\bm{x}) =  \mathcal{I} \cap \bm{x}
\end{align}
where $\mathcal{I}\cap\bm{x}$ means the subsequence containing all elements of $\bm{x}$ lying in $\mathcal{I}$, in their original order. For example, $\Pi_{[0,10]}((-1,0,1,\sqrt{117})) = (0,1)$.
\subparagraph{Restricted Point Process}

We define the restriction of our original process to the unit interval as follows:
\begin{align}
\hat{X}(D) = \lim_{i\rightarrow \infty} \Pi_{[0,1]} \left( X(D, i)\right).
\end{align}
Where we have abused the notation slightly, by applying $\Pi_{[0,1]}$ to a random variable. 
In other words, by repeatedly sampling $x_i$ from \eqref{eqn:xi} until the first $i$ such that $x_i > 1$, we obtain $\bm{x}_{<i}$ as a sample from $\hat{X}(D)$.
%
\subsection{Conditional Sampling Problem}
Roughly speaking, we wish to sample from the restricted process $\hat{X}(D)$, which is conditional on some fixed countable set of values $Z = \{z_1 < z_2 < \dots < z_r\} \subset (0,1]$ being contained within the sample. $Z$ represents the input provided by the user --- \textit{e.g.} $Z$ may represent a melodic musical fragment which we wish to harmonise using our algorithm. We introduce an additional degree of control by way of a binary indicator sequence $B = \{b_1 , b_2 , \dots , b_r\} \in \{ \true, \false\}^r$. $b_i$ denotes whether the elements are allowed to be added between $z_i$ and $z_{i+1}$, which allows sections of the musical piece (contiguous ranges of time) to be completely specified by the human. Also note that if the last element $b_r=\false$, then $z_r$ will be the final element of the generated sequence. 
Define the indicator function
\begin{align}
\label{eqn:capitald}
& \mathbbm{I}_{Z,B}(\bm{x}) = \\
& \big(Z\subseteq X\big) \text{ and } \big(\forall i: b_i=\true, \, (z_i, z_{i+1}) \cap X = \emptyset \big),
\end{align}
where $X\equiv \{ x_i \}_{i=1}^{\ell(\bm{x}) }$ is the set of elements in the sequence $\bm x$.
In words, $\mathbbm{I}_{Z,B}(\bm{x})$ is true iff $\bm x$ contains all of the elements of $Z$ while obeying the constraints indicated by $B$.
Then we may denote the random variable of interest by
\begin{align}
\hat{X}(Z,D),
\end{align}
and define it implictly by way of its p.d.f., namely
\begin{align}
\label{eqn:pxzd}
p_{\bm{x} \sim \hat{X}(Z, D)}(\bm{x}=\bm{x}')
\propto
p_{\bm{x}_Z \sim \hat{X}(D)}\left(\bm{x}_Z=\bm{x}' \, | \, \mathbbm{I}_Z(\bm{x'})\right),\quad
\end{align}
where the right hand side involves conditioning on a zero measure set, and as such should be interpreted as a \textit{regular conditional probability} \cite{regularconditionalprobability}.

\section{General Continuous Time Particle Filter}
\label{sec:pf}
\sref{Algorithm}{alg:xz} is a sequential Monte Carlo scheme for approximately sampling $\hat{X}(Z,D)$. We now demonstrate the correctness of the algorithm.

\subparagraph{Partitioned Formulation} Define the edge conditions $z_0 = -\infty$ and $z_{R+1} = \infty$, and let
\begin{align}
    r_i  = \Pi_{(z_{i-1}, z_i]}\,,
\end{align}
and
\begin{align}
    \mathcal{Z} : \, & \mathbb{R}^* \rightarrow (\mathbb{R}^*)^{R+1} \\
    & \bm{x} \mapsto \mathcal{Z}(\bm{x}) = \big( r_1(\bm{x}),r_2(\bm{x}),\dots,r_{R+1}(\bm{x}) \big). 
\end{align}
$\mathcal{Z}$ maps a sequence onto a sequence of sequences partitioned by the $z_i$. Note that the inverse of $\mathcal{Z}$ is the concatenation of the sequences. Now define
\begin{align}
    Y(Z, D) = \mathcal{Z}(\hat{X}(Z,D)),
\end{align}
where we have again abused the notation slightly by applying $\mathcal{Z}$ to a distribution. By this we mean that applying $\mathcal{Z}$ to a sample from $\hat{X}(Z,D)$ yields a sample from $Y(Z, D)$.

		\begin{algorithm}
			\caption{Our sequential Monte Carlo sampler for $\hat{X}_Z$. See \sref{Algorithm}{alg:systematicresampling} in the supplementary material for the  \textproc{SystematicResample} function. \label{alg:xz}}
			\begin{algorithmic}[1]
				\Procedure{ConditionalSample}{}\newline
				\textbf{input:} $S$, $D$, $0 < z_1 < \dots < z_R \leq 1, b_i \in \{\truet ,\falsef \}; \, \forall i$ \newline
				\textbf{output:} approximate samples $\bm{x}^{(1:S)}$ from $\hat{X}_Z$
				\For{$i$ \textbf{in} $1,2,\dots,R+1$}
				\For{$s$ \textbf{in} $\alls$}
				\If{$b_{i-1}$} {\rtext{notes allowed in $(z_{i-1}, z_i)$}}
				\While{\textbf{not} $\big( x_{\ell}\ts = z_{i}$ \textbf{or} $x_\ell\ts \geq 1\big)$}
				\State{sample $d \sim D(\bm{x}\ts)$ \rtext{time increment}}
				\State{$\bm{x}\ts \leftarrow (\bm{x}\ts, \min(z_i, x_\ell\ts + d))$ \rtext{}}
				\EndWhile
				\State{$d \leftarrow x_{\ell}\ts - x_{\ell-1}\ts$}
				\State{$w_s \leftarrow \frac{f_{D(\bm{x}_{<\ell})}(d^{(s)})}{\left(1-F_{D(\bm{x}_{<\ell})}(d^{(s)})\right)}$ \rtext{from \eqref{eq:wratio2}}} \label{line:w}
				\Else
				\If{$i \neq R+1$}{\rtext{if not the last element}}  
				\State{$\bm{x}\ts \leftarrow (\bm{x}\ts, z_i)$ \rtext{append}}
				\State{$d \leftarrow x_{\ell}\ts - x_{\ell-1}\ts$}
				\State{$ w_s \leftarrow  f_{D(\bm{x}_{<\ell})}(d^{(s)})$}
				\EndIf
				\EndIf
				\EndFor
				\If{$z_i \neq \infty$}{\rtext{if we are not finished}} \label{line:resamplingcondition}
				\State{$k_{1:S} = \textproc{SystematicResample}(w_{1:S})$
				}
				\State{$\left(\bm{x}\ts\right)_{s=1}^S \leftarrow \left(\bm{x}^{(k_s)}\right)_{s=1}^S$ \rtext{duplicate / delete}} \label{line:resampling}
				\EndIf
				\EndFor
				\State{\textbf{return} $\bm{x}_{<\ell}^{(1:S)}$}
				\EndProcedure
			\end{algorithmic}
		\end{algorithm}

Since samples from $Y(Z,D)$ are fixed length sequences (albeit with elements that are arbitrary length sequences), we can appeal to  sequential Monte Carlo logic to approximately sample from it by approximating the density of $Y(Z,D)$ by a weighted empirical distribution of $S$ \textit{particles}, namely
\begin{align}
    \tilde{p}
    (\bm{y}_{1:R+1}) 
    = 
    \sum_{s=1}^S \omega(\bm{y}_{1:R+1}\ts)
    \, \delta_{\bm{y}_{1:R+1}\ts}(\bm{y}_{1:R+1}),
\end{align}
where $\delta$ is the Dirac delta function. Note that $\bm{y}\ts_{1:R+1}$ is the $s$-th \textit{particle}, which consists of $R+1$ sequences $\bm{y}\ts_i \in \mathbb{R}^*$, for $i=1,2,\dots,R+1$. The normalised weights $\omega$ are 
\begin{align}
    \omega(\bm{y}_{1:R+1}\ts)  = 
    \frac{w(\bm{y}_{1:R+1}\ts)}{ 
    \sum_{s'=1}^S w(\bm{y}_{1:R+1}^{(s')})}.
\end{align}

\subparagraph{Proposal} We take a factorised proposal in the partitioned sequence representation,
\begin{align}
q(\bm{y}_{1:R+1}) = \prod_{i=1}^{R+1} q(\bm{y}_i|\bm{y}_{<i},Z).
\end{align}
The concrete proposal we employ, which is designed to yield a tractable particle filter scheme, is most directly understood by its sampling procedure which we provide as \sref{Algorithm}{alg:prop} of the supplementary material (see \sref{Section}{appendix:proposal}). The basic idea of the proposal is to sample forward using the recursion \eqref{eqn:xi} until such time as the relevant partition boundary $z_i$ has been crossed, at which point the element $z_i$ is included rather than the sample which crossed that value.
Mathematically, 
the proposal may be written
\begin{align}
q(\bm{y}_i^{(s)}|\bm{y}_{<i}^{(s)},Z)&=q(\bm{x}_{k+1:k+m}^{(s)}|\bm{x}_{<k+1}^{(s)},Z)\\
& = \prod_{i=1}^{m}q(\bm{x}_{k+i}^{(s)}|\bm{x}_{<k+i}^{(s)},Z),
\end{align}
where we assume there are $k$ elements $\bm{x}_{1:k}^{(s)}$ in $\bm{y}_{1:i-1}^{(s)}$, \textit{i.e.} $ \sum_{j=1}^{i-1} |\bm{y}_j^{(s)}|=k$ and $m$ elements ($\bm{x}_{k+1:k+m}^{(s)}$) in $\bm{y}_{i}^{(s)}$. 
To match the proposal sampler (\sref{Algorithm}{alg:prop} of the appendix), 
\newcommand\sssss{\hspace{-0.2mm}}
\begin{align}
 q(x_{k+1}^{(s)}|\bm{x}_{1:k}^{(s)},Z)\sssss=\sssss q(x_{k+1}^{(s)}|\bm{x}_{1:k}^{(s)},z_{i})^{b_{i}}\delta_{z_{i}}(x_{k+1}^{(s)})^{1-b_{i}},
\end{align}
(where we have let the $b_i$ act as $\{0, 1\}$ indicators), and
\newcommand\ssss{\hspace{-0.4mm}}
\begin{align}
& q(x_{k+1}^{(s)}|\bm{x}_{1:k}^{(s)},z_{i})=
\\ &f(x_{k+1}^{(s)}|\bm{x}_{1:k}^{(s)})\mathbbm{I} (x_{k+1}^{(s)}\ssss < \ssss z_{i})\ssss+\ssss\delta_{z_{i}}(x_{k+1}^{(s)})(1\ssss-\ssss F(z_i|\bm{x}_{1:k}^{(s)})),
\end{align}
where $F$ is the c.d.f. of $x_{k+m}^{(s)}$ and $\mathbbm{I}$ is the $\{0,1\}$-valued indicator function. These expressions are most easily understood by the sketches in \sref{Figure}{distribution}.


\begin{figure}
    \centering
    \begin{subfigure}[b]{0.35\textwidth}
    	\psfrag{a}{$x_{k+m-1}^{(s)}$}
    	\psfrag{b}{$z_i$}
    	\psfrag{c}{$f(x_{k+m}^{(s)})$}
    	\psfrag{d}{$F(x_{k+m}^{(s)})$}
        \includegraphics[width=\textwidth]{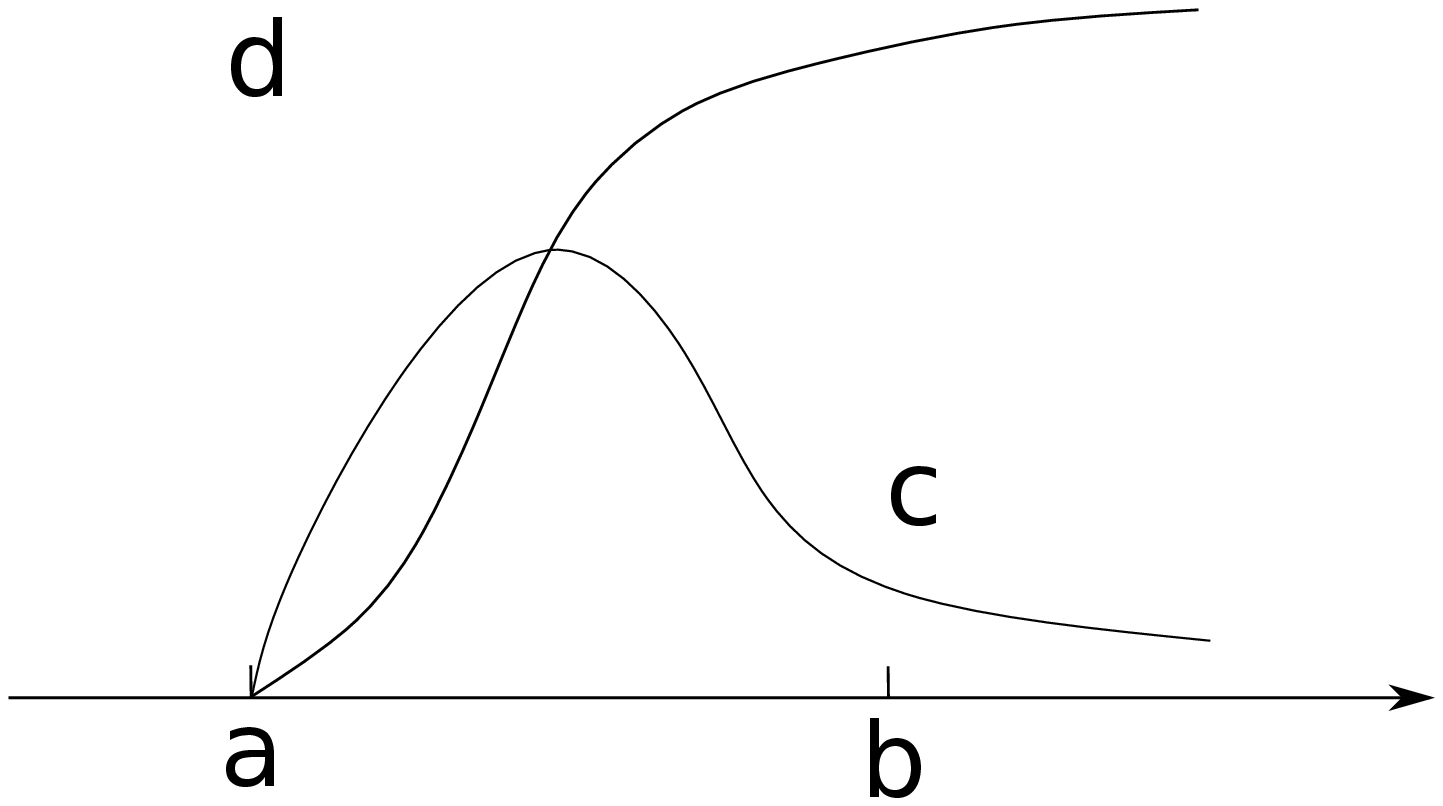}
        \caption{$f(x_{k+m}^{(s)}|\bm x_{<k+m}^{(s)})$ (of the model)}
        \label{distribution_f}
    \end{subfigure}~~~~~~
    \\
    \begin{subfigure}[b]{0.35\textwidth}
        \psfrag{e}{$x_{k+m-1}^{(s)}$}
    	\psfrag{f}{$z_i$}
    	\psfrag{g}{$Q(x_{k+m}^{(s)})$}
    	\psfrag{h}{$q(x_{k+m}^{(s)})$}
    	\psfrag{I}{$(1-F(x_{k+m}^{(s)}))\delta(x_{k+m}^{(s)})$}
    	\psfrag{1}{~}
        \includegraphics[width=\textwidth]{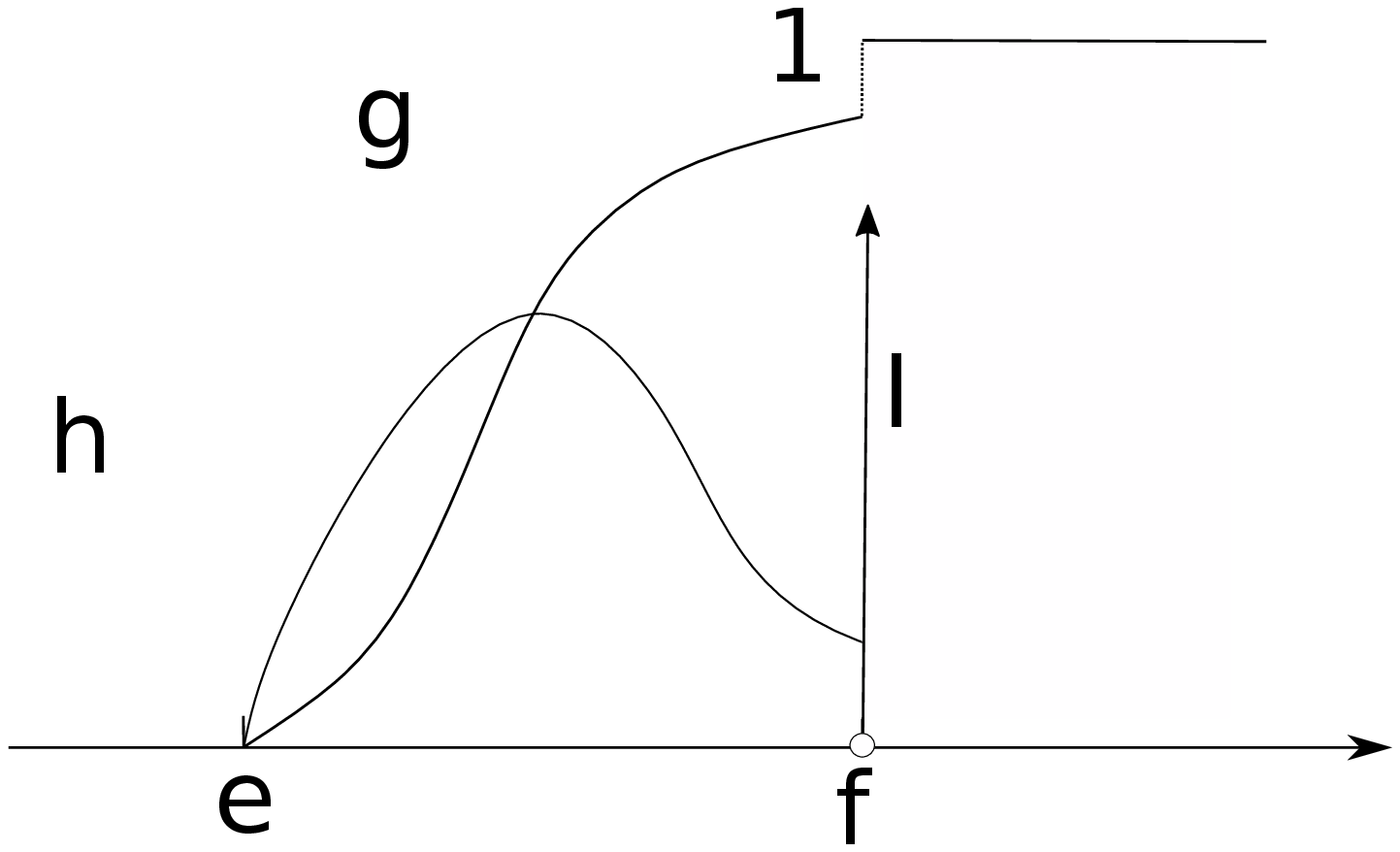}
        \caption{$q(x_{k+m}^{(s)}|\bm x_{<k+m}^{(s)})$ (of the proposal)}
        \label{distribution_q}
    \end{subfigure}~~~~~~
\caption{The model and proposal conditionals. For clarity we have omitted to notate the conditioning in the annotations of the figures.}
\label{distribution}
\end{figure}

\subparagraph{Particle Filter Resampling Weight} By sampling in the partitioned domain (of the $\bm y_i$), we have a fixed length sequence (albeit of arbitrary length sequences), and may employ standard importance sampling arguments (see \textit{e.g.} \citet{particle1993,doucet2009tutorial}) to derive the particle filtering weight of each sample,
\begin{align}
w(\bm{y}_{1:R+1}\ts) 
& = 
\frac{
    p_{\bm{y}_{1:R+1}\ts\sim Y(Z,D)}(\bm{y}_{1:R+1}\ts)
}{
    q(\bm{y}_{1:R+1}\ts)
} \\
& = 
\frac{
    \prod_{i=1}^{R+1} p(\bm{y}_i\ts | \bm{y}_{<i}\ts)
}{
    \prod_{i=1}^{R+1} q(\bm{y}_i\ts|\bm{y}_{<i}\ts)
}.
\end{align}
Note that we have omitted the subscript to $p$ in the second line. This can be decomposed recursively as
\begin{align}
w(\bm{y}_{1:i}\ts) 
& = 
\underbrace{
\frac{
    p(\bm{y}_i\ts | \bm{y}_{<i}\ts)
}{
    q(\bm{y}_i\ts|\bm{y}_{<i}\ts)
}
}_{\doteq \, w(\bm{y}_{i}\ts|\bm{y}_{<i}\ts) }
\, \times \,
\underbrace{
\frac{
    \prod_{j=1}^{i-1} p(\bm{y}_j\ts | \bm{y}_{<j}\ts)
}{
    \prod_{j=1}^{i-1} q(\bm{y}_j\ts|\bm{y}_{<j}\ts)
}
}_{\doteq \, w(\bm{y}_{<i}\ts) }. \label{eqn:wratio}
\end{align}
For $i \leq R$, assuming there are $k$ elements in $\bm{y}_{<i}$ and $m$ elements in $\bm{y}_i$ then the weight may be written
\begin{align}
w(\bm{y}_i\ts|\bm{y}_{<i}\ts) & =  \frac{
    p(\bm{y}_i\ts | \bm{y}_{<i}\ts)
}{
    q(\bm{y}_i\ts|\bm{y}_{<i}\ts)
}
= \frac{
    \prod_{j\in\mathcal{J}} p(x_j\ts|\bm{x}_{<j}\ts)
}{
    \prod_{j\in\mathcal{J}} q(x_j\ts|\bm{x}_{<j}\ts)
}\\
& =\frac{\prod_{j=k+1}^{k+m}p(x_{j}^{(s)}|\bm{x}_{<j}^{(s)})}{\prod_{j=k+1}^{k+m}q(x_{j}^{(s)}|\bm{x}_{<j}^{(s)})}
\\
&=\frac{p(x_{k+m}^{(s)}|\bm{x}_{<k+m}^{(s)}) \prod_{j=k+1}^{k+m-1} p(x_{j}^{(s)}|\bm{x}_{<j})^{(s)}}{q(x_{k+m}^{(s)}|\bm{x}_{<k+m}^{(s)}) \prod_{j=k+1}^{k+m-1} p(x_{j}^{(s)}|\bm{x}_{<j}^{(s)}) }\\
&=\frac{p(x_{k+m}^{(s)}|x_{<k+m}^{(s)})}{q(x_{k+m}^{(s)}|x_{<k+m}^{(s)})}\\
&\propto \frac{f_{D(\bm{x}_{<\ell}\ts)}(d\ts)}{1-F_{D(\bm{x}_{<\ell}\ts)}(d\ts)},
\label{eq:wratio2}
\end{align}
as per line \ref{line:w} of \sref{Algorithm}{alg:xz}. Here, $f_{D}$ and $F_{D}$ are the p.d.f. and c.d.f. of $x_{k+m}^{(s)}$ with the $d$ denoting the distance between the variable $x_{k+m}^{(s)}$ and the former note $x_{k+m-1}^{(s)}$ respectively. We have let $\bm{x}\ts=\mathcal{Z}^{-1}\bm{y}_{1:i}\ts$, recalling that $\mathcal{Z}^{-1}$ is merely the concatenation operator, and let $d\ts=x_{\ell}\ts-x_{\ell-1}\ts$. $\mathcal{J}$ are the indices into $\bm{x}\ts$ corresponding to $\bm{y}_i\ts$ (in increasing order). The final expression \eqref{eq:wratio2} follows from the preceding line because all but the final factor in the numerator and denominator cancel, due (loosely speaking) to the proposal sampling all but the last element of $\bm{y}_i\ts$ from $p$ --- a key simplification which arises from our appropriate choice of proposal distribution. The denominator of \eqref{eq:wratio2} is due to the $\min$ operation on line \ref{line:min} of \sref{Algorithm}{alg:prop} along with the definition of the c.d.f. Finally, we denote \eqref{eq:wratio2} as a proportionality by neglecting an irrelevant (and generally intractable) factor. This factor would merely renormalise the appropriate joint, to obtain the conditional on the r.h.s. of \eqref{eqn:pxzd}.

For the final step $i=R+1$ note that the $\min$ operation on line \ref{line:min} of \sref{Algorithm}{alg:prop} compares with $z_i=\infty$, thereby not altering the simple recursion of \eqref{eqn:xi}. Hence, all factors in \eqref{eqn:wratio} cancel, so that all particle weights equal one. In this case \textproc{SystematicResample} (Algorithm \ref{alg:systematicresampling}) returns the sequence $1,2,\dots,S$, leading to an identity mapping as the re-sampling step of line \ref{line:resamplingcondition} of \sref{Algorithm}{alg:xz}. Hence, for this (final) step of \sref{Algorithm}{alg:xz} we skip the re-sampling step using the condition on line \ref{line:resamplingcondition}.


\subsection{Alternative Derivation}
\label{sec:alternative}
We now sketch an alternative derivation of \sref{Algorithm}{alg:xz} (for the simplified case $b_i=\true, \forall i$) which offers an illuminating connection with the point process literature \cite{baddeleypps,danishconditionalintensity}. Assuming the process is \textit{simple} (generates no duplicate points), then we may equivalently represent samples of $\hat{X}(D)$ by way of an infinitesimally discretized indicator vector, as follows. Recall that samples of $\hat{X}(D)$ are sequences with $0 = x_0 < x_1 < \dots < x_\ell \leq 1$. Let $N$ denote the resolution of the discretization, and for $i=0,1,\dots,N-1$ define 
\begin{align}
v_i = 
\begin{cases}
1 & \text{if $x_j \in \big(i/N,(i+1)/N\big]$ for any $j$} \\
0 & \text{otherwise.}
\end{cases}
\end{align}
Consider the distribution on $\bm v$ implied by that on $\bm x$ and the above (deterministic) mapping from $\bm x$ to $\bm v$. For any $D$ (of \eqref{eqn:capitald}) there is a corresponding $g_D: \left\{0,1 \right\}^* \rightarrow [0,1]$ with
\begin{align}
\label{eqn:vbernoulli}
v_i \sim \bernoulli(g_D(\bm v_{<i})).
\end{align}
Our conditional sampling problem may now be viewed as one of sampling sequences $\bm v \in \{0,1\}^N$ conditional on $v_j = 1$ for all $j$ corresponding to the observed values $Z$. We  apply sequential Monte Carlo to this sampling problem. Let the proposal $q(v_i | \bm v_{{<i}})$ deterministically generate the value $1$ for the observed indices, and be distributed according to \eqref{eqn:vbernoulli} for the unobserved indices. By the arguments in \citet{pfrnn}, the associated particle weights are uniform (for the unobserved indices), and equal to $g_D(\bm v_{<i})$ (for the observed indices). Now, letting $k = \sum_{j=0}^{i-1} v_j$ be the index into $\bm x$ of the last point accounted for by $\bm v_{<i}$, we have
\begin{align}
g_D(v_i|\bm v_{<i}) & = \mathbb{P}[v_i=1|\bm v_{<i}] \\
& = \mathbb{P}[x_{k+1}\in \big(i/N,(i+1)/N\big] \big| \bm x_{\leq k}].
\end{align}
In the limit $N \rightarrow \infty$ with $t=i/N$ this is by definition $\lambda(t) \text{d} t$, where $\lambda$ is the conditional intensity function. It is well known that \mbox{\citep{danishconditionalintensity}}
\begin{align}
\label{eqn:conditionalintensityw}
\lambda(t) = \frac{f^*(t)}{1-F^*(t)},
\end{align}
where $f^*$ and $F^*$ are respectively the p.d.f. and c.d.f. of $x_{k+1}$ given $\bm x_{\leq k}$. We see that \eqref{eqn:conditionalintensityw} is in agreement with \eqref{eq:wratio2} (and line \ref{line:w} of \sref{Algorithm}{alg:xz}), and thereby provides an alternative intuition for \sref{Algorithm}{alg:xz}, namely that the resampling weight is proportional to the conditional intensity function of the original point process.

\section{Sequential Monte Carlo for Music}\label{SMC_music}
For simplicity, the previous section assumed a one-dimensional point process. However, multiple notes can occur at any timestep in music (e.g. chords), and as they may have arbitrary duration we need events to model their start and end. As a result, we can view music as having two dimensions, time and actions. The action space is an \textit{on} or \textit{off} action for each pitch. Although a one-dimensional sampler was constructed in the last section, an appropriate conversion can adapt it to sampling music encodings in this two-dimensional fashion. The conversion is illustrated in \sref{Figure}{convert} and explained in the next section.
 
\subsection{Modelling an Event Sequence}
We define our representation for music in two dimensions. Let a music event (which is either a note on or off event) be denoted as $e=(t(e),a(e)), t(a) \in [0, +\infty), a(e) \in[1, a_{max}]$. $t(a)$ represents the time of the event, and $a(e)$ represents the type of event, which is an action. We model music with fine temporal discretization for this conversion (each corresponding to a $2400^\text{th}$ of a musical quarter note) and 128 different MIDI pitches. We set $a_{max}=256$ with $a(e)=m, m \in[1, 128]$ representing turning on note $m$, and $a(e)=m, m \in[129, 256]$ representing turning off note $m-128$.
To reduce to one dimension we map each event to a unique integer value by setting $e = t(e)a_{max} + a(e)$. Given a one-dimensional event $e$, we can then derive its time and action by $t(e) =\lfloor(e /a_{max})\rfloor$, $a(e) = e\ \%\  a_{max}$, where $\%$ is the modulus function. 

\begin{figure}[t]
    \centering
    \begin{subfigure}[b]{0.38\textwidth}
    	\psfrag{a}{$a$}
    	\psfrag{am}{$a_{max}$}
    	\psfrag{ax}{$a(x)$}
    	\psfrag{az}{$a(z)$}
    	\psfrag{x}{$x$}
    	\psfrag{tx}{$t(x)$}
    	\psfrag{z}{$z$}
    	\psfrag{tz}{$t(z)$}
    	\psfrag{0}{0}
    	\psfrag{t}{$t$}
        \includegraphics[width=\textwidth, height=5cm]{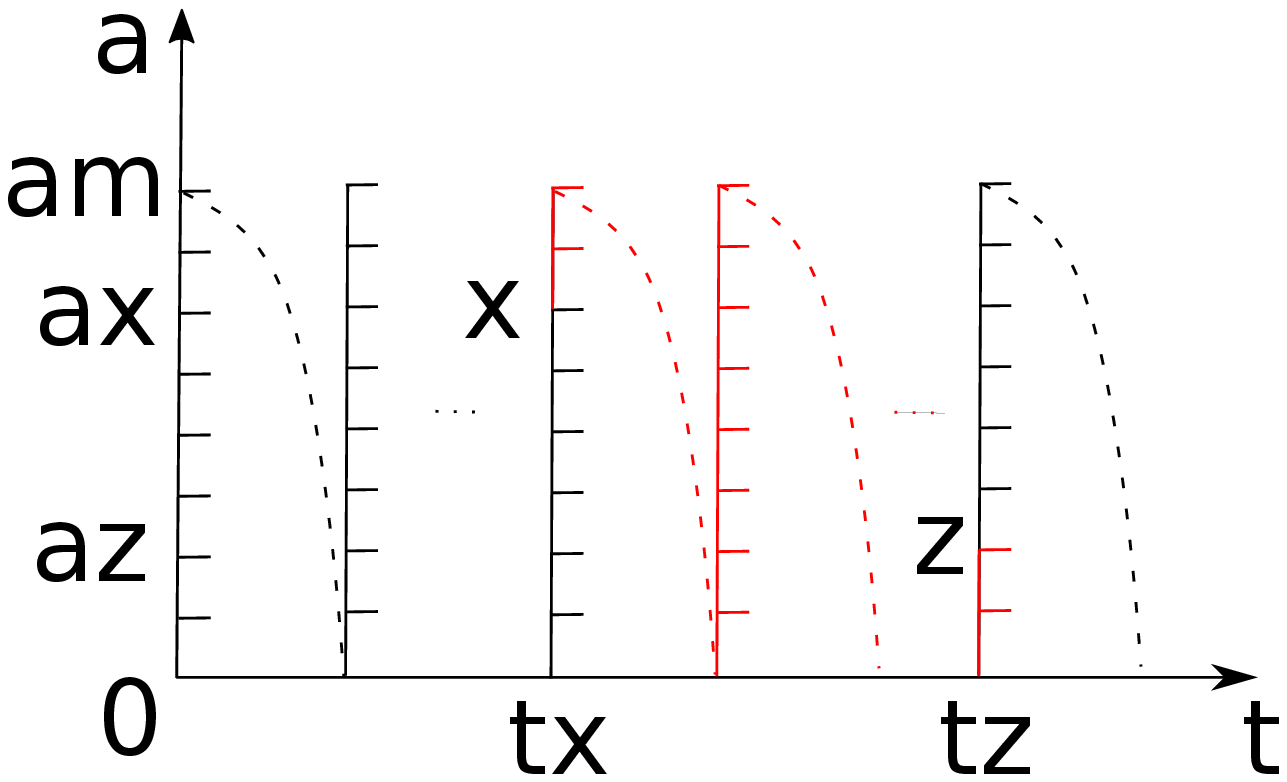}
        \caption{Two-dimensional events}
        \label{Two_D_events}
    \end{subfigure}
    \begin{subfigure}[b]{0.38\textwidth}
    	\psfrag{x}{$x$}
    	\psfrag{z}{$z$}
    	\psfrag{0}{0}
    	\psfrag{xa}{$t(x)a_{max}+a(x)$}
    	\psfrag{txa}{$t(x)a_{max}+a_{max}$}
    	\psfrag{tam}{$t(z)a_{max}$}
    	\psfrag{za}{$t(z)a_{max}+a(z)$}
        \includegraphics[width=\textwidth, height=5cm]{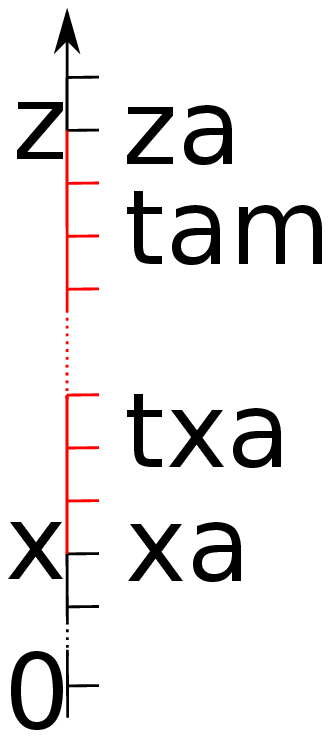}
        \caption{Unrolled one-dimensional events}
        \label{one_D_events}
    \end{subfigure}
\caption{A sketch of the reduction from two- to one-dimensional event domains.}
\label{convert}
\end{figure}

\subsection{Modelling with an RNN}\label{rnn_modelling}
We model the distribution of music using a Recurrent Neural Network (RNN) based model with a music representation inspired by the \textit{Performance RNN} \citep{performancernn}, expanding the action space to include different \textit{parts} (MIDI channels). Each symbol modeled by the RNN is either an action (a musical note being turned on or off in a specific part) or a shift in time between music events. Notes turned on prior to the shift are kept on during the time shift, and there is be silence otherwise. 

\subparagraph{Resolving Ambiguity}
We improve slightly on \citet{performancernn} by ensuring that our representation for a given musical sequence is unique --- or roughly speaking that there is a bijection between MIDI data and what the RNN sees. This is achieved by simply enforcing that shifts not follow shifts, and at a given time, actions can only occur in ascending order of enumeration. 
\begin{table*}
\centering
\begin{tabular}{|c|c|c|}\Xhline{2\arrayrulewidth}
 \backslashbox[10mm]{Variable~~~~~~~}{$t$~~~~}
  &\makebox[7em]{$t(z)=t(x_{l-1})$} &\makebox[8em]{$t(z)>t(x_{l-1})$}\\ \Xhline{2\arrayrulewidth}
$x_l \sim \bm{x}_{<l}$ & sample to $a(z)$ & shift to time $t(z)$, then sample to action $a(z)$\\
\hline
$f_{\bm{x}_{<l}}(x_l)$ & $f_{\bm{x}_{<l}}(a(z))$ & $f_{\bm{x}_{<l}}(t(z)-t_x) \cdot f_{z}(a(z))$\\
\hline
$F_{\bm{x}_{<l}}(x_l)$ & $\sum_{a(x)}^{a(z)} {f_{\bm{x}_{<l}}(a)}$ & \tabincell{c}{$\sum_{a_x}^{a_z} {f_{\bm{x}_{<l}}(a)}+\sum_{1}^{t(z)-t(x)-1} {f_{\bm{x}_{<l}}(\Delta t)}+$ \\ ${f_{\bm{x}_{<l}}(t(z)) \cdot \sum_{1}^{a(z)} {f_z(a)}}$}\\
\hline
\end{tabular}
\caption{Summary of the key quantities required by \sref{Algorithm}{alg:xz}. Here, $f_z$ (respectively $F_z$) denotes the conditional p.d.f. (c.d.f.) of the next step, given that the particle has reached $t(z)$. \label{notation}}
\end{table*}
%
To this end, we define a constant mask matrix $M$ which has the following structure:
\[
M=
\begin{gathered}
\begingroup%
  \makeatletter%
  \providecommand\color[2][]{%
    \errmessage{(Inkscape) Color is used for the text in Inkscape, but the package 'color.sty' is not loaded}%
    \renewcommand\color[2][]{}%
  }%
  \providecommand\transparent[1]{%
    \errmessage{(Inkscape) Transparency is used (non-zero) for the text in Inkscape, but the package 'transparent.sty' is not loaded}%
    \renewcommand\transparent[1]{}%
  }%
  \providecommand\rotatebox[2]{#2}%
  \newcommand*\fsize{\dimexpr\f@size pt\relax}%
  \newcommand*\lineheight[1]{\fontsize{\fsize}{#1\fsize}\selectfont}%
  \ifx\svgwidth\undefined%
    \setlength{\unitlength}{177.42373849bp}%
    \ifx\svgscale\undefined%
      \relax%
    \else%
      \setlength{\unitlength}{\unitlength * \real{\svgscale}}%
    \fi%
  \else%
    \setlength{\unitlength}{\svgwidth}%
  \fi%
  \global\let\svgwidth\undefined%
  \global\let\svgscale\undefined%
  \makeatother%
  \begin{picture}(1,0.67550444)%
    \lineheight{1}%
    \setlength\tabcolsep{0pt}%
    \put(0,0){\includegraphics[width=\unitlength]{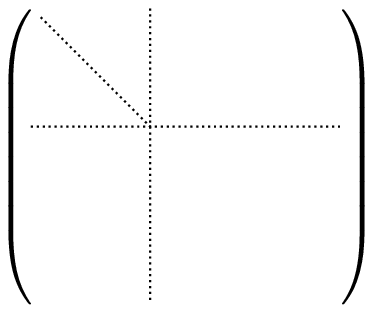}}%
    \put(0.69482542,0.59686246){\color[rgb]{0,0,0}\makebox(0,0)[t]{\lineheight{1.25}\smash{\begin{tabular}[t]{c}$a_{max} + s_{max}$\end{tabular}}}}%
    \put(0.35164321,0.59686247){\color[rgb]{0,0,0}\makebox(0,0)[t]{\lineheight{1.25}\smash{\begin{tabular}[t]{c}$a_{max}$\end{tabular}}}}%
    \put(0.76113979,0.38213382){\color[rgb]{0,0,0}\makebox(0,0)[t]{\lineheight{1.25}\smash{\begin{tabular}[t]{c}$a_{max}$\end{tabular}}}}%
    \put(0.84190597,0.09579442){\color[rgb]{0,0,0}\makebox(0,0)[t]{\lineheight{1.25}\smash{\begin{tabular}[t]{c}$a_{max} + s_{max}$\end{tabular}}}}%
    \put(0.27377874,0.50179493){\color[rgb]{0,0,0}\makebox(0,0)[t]{\lineheight{1.25}\smash{\begin{tabular}[t]{c}$-\infty$\end{tabular}}}}%
    \put(0.20296179,0.42731061){\color[rgb]{0,0,0}\makebox(0,0)[t]{\lineheight{1.25}\smash{\begin{tabular}[t]{c}$0$\end{tabular}}}}%
    \put(0.23920331,0.22856949){\color[rgb]{0,0,0}\makebox(0,0)[t]{\lineheight{1.25}\smash{\begin{tabular}[t]{c}$0$\end{tabular}}}}%
    \put(0.49896258,0.4620761){\color[rgb]{0,0,0}\makebox(0,0)[t]{\lineheight{1.25}\smash{\begin{tabular}[t]{c}$0$\end{tabular}}}}%
    \put(0.4972983,0.22926844){\color[rgb]{0,0,0}\makebox(0,0)[t]{\lineheight{1.25}\smash{\begin{tabular}[t]{c}$-\infty$\end{tabular}}}}%
  \end{picture}%
\endgroup%

\end{gathered}
\]
and, denoting the output layer before the softmax layer as $y_l=f_{\bm{x}_{<l}}(x_l)$, we then replace $y_l$ with
\begin{align}
z_l=y_l+M(a(x_{l-1})),
\label{eq:mask}
\end{align}
where $M(a(x_{l-1}))$ denotes the $a(x_{l-1})$-th column of $M$. 


\subparagraph{Unrolled Probabilities in terms of Event Probabilities}
To apply \sref{Algorithm}{alg:xz}, we must write the required quantities as they pertain to the \textit{unrolled} music representation (that is, the bottom of \sref{Figure}{convert}) in terms of the probabilities available in the original, unrolled representation (the top of \sref{Figure}{convert}).
Suppose the current event step in the sequence is $x$, then the probability mass function (PMF) vector of the next potential step $f_x$ is obtained through a softmax layer that is divided into two parts: $f_x=\big[f_x(a): a \in [1,a_{max}], f_x(\Delta t): \Delta t \in [1,s_{max}]\big]$.
$f_x(a)$ gives the probabilities for different actions at the current time step $t(x)$, and $f_x(\Delta t)$ gives the probabilities for different shift times. Each element of $f_x(\Delta t)$ is the sum of the probabilities of all the actions at $t(x)+\Delta t$. The key variables are summarized in \sref{Table}{notation}, which provides the basis for computing the weights of the particles in \eqref{eq:wratio2} as required in \sref{Algorithm}{alg:xz}.


\section{Results}
\label{sec:results}
Algorithm \ref{alg:xz} was used to generate music samples using the conversion for music from Section \ref{SMC_music}. 
An RNN trained on the Symbolic Music Data Version 1.0 (SMD-V1) \cite{SMD} was applied for the sampling distribution $D$  in Algorithm \ref{alg:xz}. SMD-V1 consists of 5 different music datasets, and the validation sets of two datasets (\texttt{PMD} and \texttt{NOT}) were used for testing. The pieces in these two datasets are separated into two parts (MIDI channels). These are labelled as Part 0 and Part 1. Using this combined validation set, the following 6 categories of samples were generated:

\begin{description}
	\item[GroundTruthSamples:] 20s samples taken at random directly from samples in the validation sets with both parts intact. Labelled as GT in the figures.
	\item[ConditionedPF-$\{s\}$:] 40s samples were taken at random from the test set of the dataset. The RNN was conditioned on he first 20s, and then Algorithm \ref{alg:xz} was run with this conditioned model as distribution $D$ and the notes in Part 0 of the second 20s as the constraints. $s$ particles were used in the algorithm. Labelled as PF1, PF2, PF3 in order of ascending $s\in\{30,100,300\}$ in the figures.
	\item[ConditionedBeamSearch-${b}$-${f}$:] Used the same setup as ConditionedPF-$\{s\}$ but with a beam search based algorithm in place of Algorithm \ref{alg:xz}. The beam search utilizes Algorithm \ref{alg:prop} while retaining multiple beams. In the loop (lines 5-9) of Algorithm \ref{alg:prop} it samples $b\times f$ trajectories at random (specifically $b$ samples from each the previously kept $f$ trajectories), and then keeps the top $f$. The ranking is based on maximising log probabilities of the current (incomplete) trajectories. Labelled as BS1, BS2  in order of ascending $f$ in the figures.
\end{description}

\subparagraph{Parameter choice} The parameters ($s, b$ and $f$) were chosen to give an interesting comparison in the number of trajectories, amount of memory and time taken. In particular, BS1 has as many trajectories as PF3, while BS2 has vastly more than the other algorithms, giving it an unfair advantage. These performance metrics are detailed in the Supplementary Material, Section \ref{sec:performance}.

To measure the quality of the resulting samples, in the following sections we investigate two aspects of the output: the probability of our samples under the distribution of our underlying RNN, and the results of human listening tests.

\subsection{Probability of the Generated Sequences}
\begin{figure}
	\centering
	\includegraphics[width=\columnwidth]{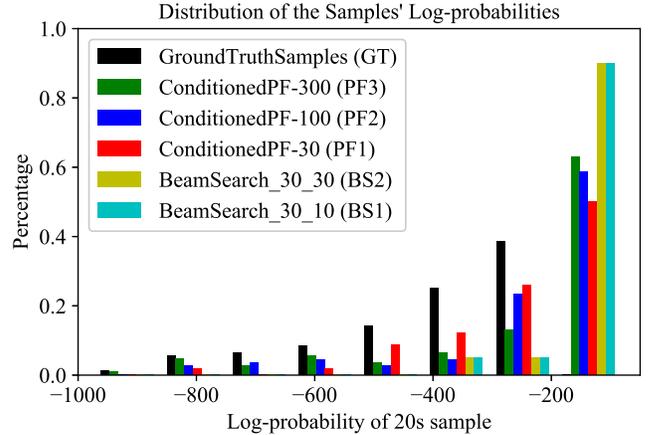}
	\caption{Histograms of log-probabilities.\label{LogProbs}}
\end{figure}

From \eqref{factorised}, we evaluate the log probability of a sample by
\begin{align*}
	\log p(\bm{x}_{1:i}) & = \sum_{k=1}^{i}\log f(x_{k}|\bm{x}_{<k}).
\end{align*}
where $f$ is our model distribution given by the RNN. 
A histogram of the log probabilities of the samples from each category is shown in Figure \ref{LogProbs}. 
BeamSearch's distribution of log probabilities is much higher than ConditionedPF's, with larger proportions at $-100$ and smaller range (-400 to -50 compared to -800 to -50 for ConditionedPF). Furthermore, as the number of maintained trajectories increases, the distributions for both algorithms become thinner and sharper, overall being more weighted towards higher log probabilities. 
This is the difference between the beam search based approach, which is actively maximising the whole subsequence between constraints, and Algorithm \ref{alg:xz} which is resampling based on the probability of constrained events only. On the other hand, the GroundTruthSamples have a range from $-1000$ to $-200$, and a mode at $-300$, and which is most similar to the ConditionedPF. The distribution of GroundTruthSamples being very different to the BeamSearch is the expected result of actively maximising log probabilities rather than generating samples from the conditional distribution of interest as in the particle filter method. This also does not allow for the diversity typical of human music \citep{pfrnn} and makes the samples rather bland and repetitive as we demonstrate in the next section.

\subsection{Listening Tests}

\begin{table}
	\centering
	{\small
		\begin{tabular}{c|c c c c c c}\Xhline{2\arrayrulewidth}
			& BS1 & BS2 & PF1 & PF2 & PF3 & GT\\ \Xhline{2\arrayrulewidth}
			BS1       &  -   &  0.51 &  0.51 &  0.54 &  0.49 &  0.39 \\
			BS2       & 0.49 &   -   &  0.52 &  0.44 &  0.42 &  0.44 \\
			PF1       & 0.49 &  0.48 &   -   &  0.54 &  0.52 &  0.46 \\
			PF2       & 0.46 &  0.56 &  0.46 &   -   &  0.59 &  0.40 \\
			PF3       & 0.51 &  0.58 &  0.48 &  0.41 &   -   &  0.48 \\
			GT        & 0.61 &  0.56 &  0.54 &  0.60 &  0.52 &   -  
			\\\hline
		\end{tabular}
	}
	\caption{Row beats column ratio for the ``Better'' test. \label{better_table}}
\end{table}
\begin{table}
	\centering
	{\small
		\begin{tabular}{c|c c c c c c}\Xhline{2\arrayrulewidth}
			& BS1 & BS2 & PF1 & PF2 & PF3 & GT\\ \Xhline{2\arrayrulewidth}
			BS1       &  -   &  0.42 &  0.33 &  0.41 &  0.33 &  0.28 \\
			BS2       & 0.58 &   -   &  0.35 &  0.37 &  0.37 &  0.30 \\
			PF1       & 0.67 &  0.65 &   -   &  0.46 &  0.40 &  0.40 \\
			PF2       & 0.59 &  0.63 &  0.54 &   -   &  0.45 &  0.34 \\
			PF3       & 0.67 &  0.63 &  0.60 &  0.55 &   -   &  0.43 \\
			GT        & 0.72 &  0.70 &  0.60 &  0.66 &  0.57 &   -  
			\\\hline
		\end{tabular}
	}
	\caption{Row beats column ratio for the ``Interesting'' test. 
		\label{interesting_table}}
\end{table}

\subparagraph{Setup} A listening test similar to \citet{maestro} was used for the six categories. Sixteen 20s samples from each category were taken for comparative listening tests between each pair from different categories. After listening to the two samples (labelled `A' and `B'), the listeners were asked ``Which sounds better overall?'' and ``Which sounds more interesting?''. For both questions, the listeners were given the choice `A' or `B'. Both the order of the tests and the order of the two samples within each test were randomised. Every individual test was also given to 3 different listeners, making a total of $3\times (^6_2) \times 16^2=11,520$ listening tests.

\begin{figure}
	\centering
	\includegraphics[width=\columnwidth]{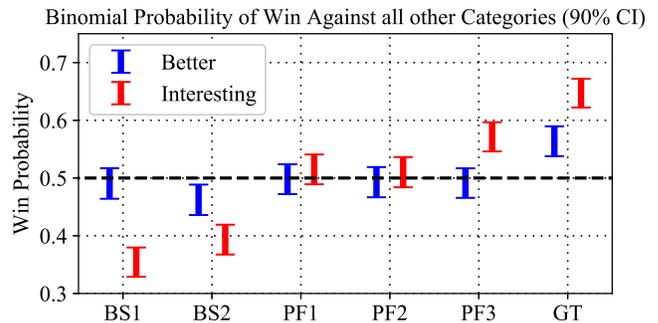}
	\caption{Listening test wins against all other categories. \label{votes}}
\end{figure}

\subparagraph{Overall Wins} Overall win proportions can be seen in Figure \ref{votes}. For the `Better' listening test, all categories did well (around 50\%) while the GroundTruth category achieved 56\%. This shows that while people clearly percceived the ground truth samples as superior, samples from each category were convincing enough to win almost half the time. On the other hand, for the `Interesting' test, BeamSearch performed significantly lower than ConditionedPF (at almost 40\% wins compares to 50-60\% wins), with ConditionedPF-300 performing very well (57\% wins) and GroundTruth clearly peforming the best (65\% wins). This suggests BeamSearch acts as expected of a greedy algorithm: it seeks to optimise its likelihood (which intuitively would make it more likely to win the `Better' test), and therefore, it plays very monotonous but `safe' pieces, leading to uninteresting music samples. However, ConditionedPF is able to sample more `risky' and interesting pieces while still maintaining high likelihoods, thus doing well on both tests, especially with higher particle sizes.

\begin{figure}
	\centering
	\includegraphics[width=\columnwidth]{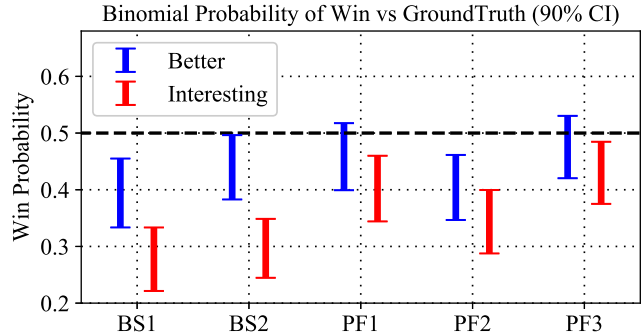}
	\caption{Listening test wins \textit{vs.} GroundTruth.\label{binom}}
\end{figure}

\subparagraph{Comparisons with the Ground Truth} Direct comparisons for each category versus the GroundTruth samples can be seen in Figure \ref{binom}. There is a larger difference between the BeamSearch and the ConditionedPF on the `Better' test, while the BeamSearch samples compare poorly with the GroundTruth. The ConditionedPF with 300 particles performs the best, nearly winning half of the tests (47.5\%). Likewise, the BeamSearch rates poorly on the `Interesting' test, and again ConditionedPF-300 performs the best (43\%). Detailed comparisons between each pair of categories can be seen in Tables \ref{better_table} and \ref{interesting_table}.


\section{Conclusion}
\label{sec:conclusion}
We have presented a novel sequential Monte Carlo algorithm for continuous time point processes that are conditioned on a subset of events. By a suitable choice of proposal distribution, we obtained Monte Carlo resampling weights that are computationally tractable and simple to compute.  We demonstrated how this general approximate sampling algorithm may be adapted to symbolic music generation, by letting the user specify a subset of musical events to be included in the piece, along with controls on which parts of the piece to hold fixed. Our results show that the proposed scheme produces more interesting samples compared to a beam search baseline, without sacrificing musical quality as perceived by human test subjects. The samples produced by our novel approach sound more like human music,  perform better under two listening test metrics, and have a  distribution which more closely resembles human compositions when evaluated using our generative model for music.

\bibliographystyle{aaai}
\bibliography{walder}

\clearpage

\appendix
\section{Supplementary Material for ``\textit{\thetitle}''}
\subsection{Proposal Sampler}
\label{appendix:proposal}
We define the proposal $q(\bm{y}_i|\bm{y}_{<i})$ procedurally as \sref{Algorithm}{alg:prop}. The proposal samples appropriately from the interval $(z_{i-1}, z_i]$ by applying the recursion \eqref{eqn:xi}, but with a clipping operation on line \ref{line:min} in order to include $z_i$ in the proposed sequence with probability one. Once $z_i$ is included the algorithm terminates. Note that the edge conditions at lines \ref{line:edge1} and \ref{line:edge2} of \sref{Algorithm}{alg:prop} vanish in the context of the complete \sref{Algorithm}{alg:xz}, which is more elegant than the proposal distribution.
%
%
\begin{algorithm}[h!]
	\caption{A sampler for the proposal $q(\bm{y}_i|\bm{y}_{<i})$ \label{alg:prop}}
	\begin{algorithmic}[1]
		\Procedure{Proposal\label{peoposal_procedure}}{} 
		\newline
		\textbf{input:} $i, \bm{y}_{<i}$, $D$, $0 < z_1 < \dots < z_R \leq 1$, $b_{i-1}.$
		\newline
		\textbf{output:} $\bm{y}_i \sim q(\bm{y}_i|\bm{y}_{<i})$
		\If{$i=1$} \label{line:edge1}
		\State{$x \leftarrow 0;\quad\bm{y}_i\ts \leftarrow (0)$ 
		}
		\Else
		\State{$x \leftarrow (\bm{y}_{i-1})_\ell 
		;\quad \bm{y}_i\ts \leftarrow ()$ 
		}
		\EndIf
		\State{$x \leftarrow (\bm{y}_{i-1})_\ell$ \rtext{last value of previous subsequence}}
		\State{$\bm{y}_i\ts \leftarrow ()$ \rtext{length zero sequence}}
		\If{$b_{i-1}$} {\rtext{notes allowed in $(z_{i-1}, z_i)$}}
		\While{$\big( x< z_i \big)$}
		\State{$\bm{x}\ts \leftarrow \mathcal{Z}^{-1}(\bm{y}_{1:i}\ts)$ \rtext{$\mathcal{Z}^{-1}$ concatenates}}
		\State{sample $d \sim D(\bm{x}\ts)$}
		\State{$\bm{y}_i \leftarrow (\bm{y_i}, \min(z_i, x+d))$ \rtext{append}} \label{line:min}
		\State{$x \leftarrow (\bm{y}_i)_\ell$}
		\EndWhile
		\Else
		\State{$\bm{y}_i \leftarrow (\bm{y_i},z_i)$ \rtext{append}}
		\EndIf
		\State \Return{$\bm{y}_i$}
		\If{$x > 1$} \label{line:edge2}
		\State \Return{$(\bm{y}_i)_{<\ell}$}
		\Else
		\State \Return{$\bm{y}_i$}
		\EndIf
		\EndProcedure
	\end{algorithmic}
\end{algorithm}

%
\begin{figure}[h!]
		\begin{algorithm}[H]
			\caption{Systematic Resampling\\ \protect\citep{systematicresampling}.}
			\label{alg:systematicresampling}
			\begin{algorithmic}[1]
				\Procedure{SystematicResample}{}\newline
				\textbf{input:} $w_{1:S}$ \newline
				\textbf{output:} $k_{1:S}$
				\State{$\omega_{1:S} \leftarrow w_{1:S} / \sum_{s=1}^S w_{s}$}
				\State{sample $u \sim \textproc{Uniform}([0,1])$}
				\State{$\bar{u}\leftarrow u / S$}
				\State{$j\leftarrow 1$}
				\State{$S_\omega \leftarrow \omega_1$}
				\For{$l$ \textbf{in} $\alls$}
				\While{$S_\omega < \bar{u}$}
				\State{$j \leftarrow j + 1$}
				\State{$S_\omega \leftarrow S_\omega + \omega_j$}
				\EndWhile
				\State{$k_l\leftarrow j$}
				\State{$\bar{u} \leftarrow \bar{u}+1/S$}
				\EndFor
				\State\Return{$k_{1:S}$}
				\EndProcedure
			\end{algorithmic}
		\end{algorithm}
\end{figure}

%
\subsection{Music Samples and Listening Tests}
The samples used for the listening test can be found at
\begin{verbatim}
http://bit.ly/GenMusicSamples
\end{verbatim}
It is separated into 6 folders, one for each category, and within each there are 16 music samples rendered into mp3 format.

Furthermore, in our supplementary material we provide four selected music samples from each category, which correspond to the four most interesting pieces as determined by the ``interesting'' listening tests.

The output music events from our algorithm were converted into MIDI (our representation gives a unique representation to MIDI). These were then synthesized using a custom SoundFont based on piano, and then converted to mp3. No other adjustments were made.

The listening tests were conducted on Amazon Turk. Each listening test was set as a different task, and there were 3 assignments per task (this means 3 different workers were required to complete each listening test). 767 workers completed our 11,520 listening tests, taking a median of 69 seconds to complete.


\subsection{Comparison of Algorithms}
\label{sec:performance}

\begin{table*}[h]
	\centering
	\begin{tabular}{|c|cccc|}
	\Xhline{2\arrayrulewidth}
		Algorithm & Trajectories & Memory & Time Taken (s) & Survived (\%)\\ \Xhline{2\arrayrulewidth}
		ConditionedPF-30 (PF1) 	& 30	&	30		&	157		&	0.44	\\
		ConditionedPF-100 (PF2)	& 100	&	100		&	543		&	0.83	\\
		ConditionedPF-300 (PF3)	& 300	&	300		&	1829	&	0.84	\\
		BeamSearch-30-10 (BS1)	& 300	&	30		&	1057	&	0.73	\\
		BeamSearch-30-30 (BS2)	& 900	&	30		&	3344	&	0.74	\\\hline
	\end{tabular}
	\caption{Results for each generated category}\label{results}
\end{table*}

Table \ref{results} shows a summary of the performance of the algorithms mentioned in Section \ref{sec:results}. Trajectories indicates how many different subsequences between each two constraints are sampled, $b\times f$ for BeamSeach and $s$ for ConditionedPF. It can be considered as a measure of the search space of the algorithm. Memory is the number of trajectories kept when a constraint is reached, $f$ for BeamSearch and $s$ for ConditionedPF. Thus it is the number promising sequences the algorithm keeps. Time taken is the average time for each algorithm to make a sample. The parameters $b$, $f$ and $s$ for the last 5 categories were chosen to give a good comparison between BeamSearch and ConditionedPF with similar Trajectory and Memory values, within a reasonable amount of time (an hour). Experiments were run on an i7-6500 Intel CPU.\\

Each algorithm also has a survival percentage. This is the proportion of times the algorithm returned a complete conditioned music sample rather than failing. A failure happens when there is no choice but to sample a zero probability event on all trajectories. This is caused by the forced event of the constraints (the $\min$ operator returning $z_i$ on line 8 in Algorithm \ref{alg:prop} or line 7 in Algorithm \ref{alg:xz}) being too unnatural to make any other continuations likely. Thus the survival rate can be considered as how likely the algorithm will avoid impossible sequences and find probable sequences subject to the constraints.\\

Time taken is most dependent on the number of trajectories and then on memory. The BeamSearch algorithms have the most trajectories and take $\approx$ 20mins and $\approx$ 1 hour respectively, while having a survival rate of at most 74\%. However, ConditionedPF-100 takes just under 10mins with a much higher rate of survival. It is interesting to note that changing from 30 particles to 100 particles in ConditionedPF increases the survival rate greatly, from 44\% to 83\%. Having 300 particles only increased the survival rate marginally, to 84\%.

\end{document}